\documentclass[twocolumn, twocolappendix]{emulateapj}
\usepackage{natbib}
\usepackage{graphicx}
\usepackage{CJK}
\usepackage[version=4]{mhchem}
\usepackage{amsmath}
\usepackage{subfigure}
\usepackage{multirow}
\usepackage{wrapfig}
\usepackage{color,xcolor}
\usepackage{longtable}

\definecolor{ao}{RGB}{0,103,192}
\definecolor{gunen}{RGB}{174,0,46}

\usepackage[colorlinks=true,citecolor=ao,linkcolor=gunen,urlcolor=ao]{hyperref}

\graphicspath{{./}{figures/}}

\received{-}
\accepted{2019-11-23}
\submitted{The Astrophysical Journal Supplement Series, 246:12 (14pp), 2020 January}

\shorttitle{Compact Groups of Galaxies in SDSS and LAMOST: I. The Catalogs}
\shortauthors{Zheng et al.}

\begin{document}
\begin{CJK*}{UTF8}{gbsn}

\title{Compact Groups of Galaxies in Sloan Digital Sky Survey and LAMOST Spectral Survey: I. The Catalogues\footnote{}}

%\correspondingauthor{Shi-Yin Shen}
%\email{ssy@shao.ac.cn}

\author{
Yun-Liang Zheng (郑云亮)\altaffilmark{1,2},
Shi-Yin Shen (沈世银)\altaffilmark{1,3}}

\altaffiltext{1}{Key Laboratory for Research in Galaxies and Cosmology, Shanghai Astronomical Observatory, Chinese Academy of Sciences, 80 Nandan Road, Shanghai, China, 200030, ssy@shao.ac.cn}
\altaffiltext{2}{University of the Chinese Academy of Science, No.19A Yuquan Road, Beijing, China, 100049}
\altaffiltext{3}{Key Lab for Astrophysics, Shanghai, China, 200034}

\begin{abstract}
A compact group (CG) is a kind of special galaxy system where the galaxy members are separated at the distances of the order of galaxy size.  The strong interaction between the galaxy members makes CGs ideal labs for studying the environmental effects on galaxy evolution.  The traditional photometric selection algorithm biases against the CG candidates at low redshifts, while the spectroscopic identification technique is affected by the spectroscopic incompleteness of sample galaxies and typically biases against the high redshift candidates. In this study, we combine these two methods and select  CGs in the main galaxy sample of the Sloan Digital Sky Survey,  where we also have taken the advantages of the complementary redshift measurements from the LAMOST spectral and GAMA surveys. We have obtained the largest and most complete CG samples to date. Our samples include 6,144 CGs and 8,022 CG candidates, which are unique in the studies of the  nature of the CGs and the evolution of the galaxies inside.
\end{abstract}

\keywords{catalogs -- galaxies: groups: general -- surveys}

\section{Introduction} \label{sec:intro}
Galaxies are the building blocks of the visible universe, inhabiting a variety of environments from fields to galaxy clusters. Observations show that over half of the galaxies are located in group systems which have members from a few to dozens (\citealp{1982ApJ...257..423H},\citealp{2007ApJ...671..153Y},\citealp{2012AA...540A.106T}). A compact Group of galaxies (hereafter CG) is an extreme case of groups which contains a few member galaxies separated by projected distance of the order of galaxy size. CG is believed to represent an environment where tidal interaction (\citealp{2007AJ....133.2630C}), harassment (\citealp{1998ApJ...495..139M}) and galaxy merging (\citealp{1989Natur.338..123B}) are much more active than in a normal/loose group of galaxies. The high density and low velocity dispersion make CG to be an ideal laboratory for studying galaxy interaction and the merging process. Indeed, both simulations \citep{2009MNRAS.392.1141B} and observations (\citealp{2004AJ....127.1811L}, \citealp{2008AA...484..355D}) have shown that the fraction of early-type galaxies in CGs is significantly higher than the counterparts in normal groups and fields. \citet{2012AA...543A.119C} found that galaxies in CGs have systematically larger concentration index and higher surface brightness and further concluded that the star-forming galaxies are more likely to be quenched in the CG environment. 

In the aspect of theoretical modeling, we are not very clear about the formation channel of CGs. Given the short merging timescale of galaxies in the CG environment, the occurrence of CGs are tightly correlated with their formation process and effective lifetime. \citet{1994AJ....107..868D} proposed that CGs are embedded in larger systems and so that they can be constantly replenished from neighboring galaxies. \citet{2016ApJS..225...23S} found that the occurrence of CGs has no significant change with redshift and thus suggest a long life span of CGs. Also, CGs were once believed to be the progenitors of fossil groups \citep{1994Natur.369..462P}, but later \citet{2017ApJ...840...58F} suggested that although CGs are systematically younger than fossil groups, their evolutionary paths are different and the evolution of CGs will not lead to the formation of the fossil groups.

\begin{figure}
\begin{center}
	\includegraphics[width=\columnwidth]{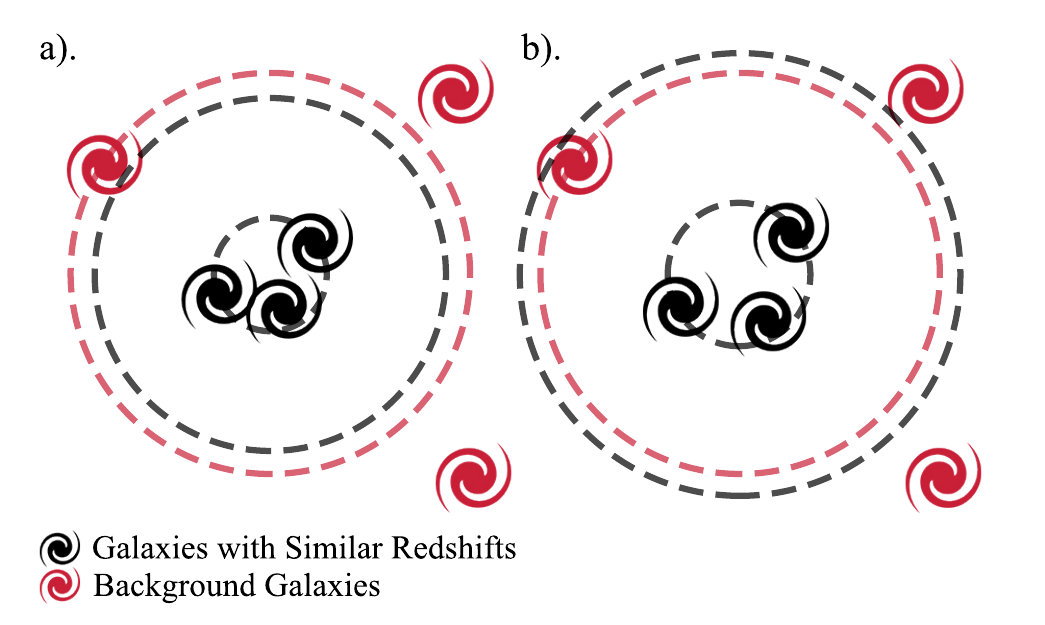}
    \caption{Assumed compact group at different redshifts. The distant one (left panel) has smaller angular separation and is more likely to satisfy $\theta_{n} \geq 3$ $\theta_{G}$, while the nearby one (right panel) is more likely to violate the $\theta_{n} \geq 3\theta_{G}$ criterion.} \label{fig:bias}
\end{center}
\end{figure}

To further shed light on the formation and evolution path of the CGs, a large  sample of CGs with redshift measurements and well-defined selection effects are required, which is not a easy task even in these days with many modern galaxy surveys.  Historically, CG samples were constructed based on photometric surveys. \citet{1982ApJ...255..382H} first introduced a set of criteria based on photometric information and identified 100 CGs (HCGs) from the Palomar Observatory Sky Survey. The three Hickson's criteria are:

\begin{enumerate}
\setlength{\itemsep}{-0.7ex}
\item Richness: $ N\left(\Delta\,m \leq 3\right) \geq 4$ 
\item Isolation: $ \theta_{n} \left(\Delta\,m \leq 3\right) \geq 3 \theta_{G}$ 
\item Compactness: $ \mu \leq 26.0$ mag\,arcsec$^{-2}$ 
\end{enumerate}

where $N\left(\Delta m \leq 3\right)$ is the total number of galaxies within 3 magnitudes of the brightest member, $\mu$ is the effective surface brightness averaged over the smallest enclosing circle with angular radius $\theta_{G}$, $\theta_{n} \left(\Delta\,m \leq 3\right)$ is the angular radius of the largest concentric circle that contains no external galaxies within the same magnitude range. With follow-up spectroscopic measurements, \citet{1992ApJ...399..353H} further removed 8 HCGs which contain fewer than three accordant members. This formed a tradition of the CG selection: searching the CG candidates that satisfy all the Hickson's criteria in the first step, then discarding the groups containing interlopers according to their redshifts. Thereafter, many CG catalogs have been constructed following this methodology: from the COSMOS-UKST Southern Galaxy Catalog \citep{2002AJ....124.2471I}; from the DPOSS Catalog \citep{2003AJ....125.1660I}; from the Two Micron All Sky Survey (2MASS) extended source catalog \citep{2012MNRAS.426..296D} and from the Sloan Digital Sky (SDSS) galaxy catalog (\citealp{2004AJ....127.1811L}, \citealp{2009MNRAS.395..255M}, \citealp{2015JKAS...48..381S}). This kind of post-HCG  procedure is very efficient in identifying real CGs, which however inherits a bias against nearby CGs in its photometric selection. For the nearby CGs, because of its large angular diameter,the isolation criteria is more likely to be broken given the background galaxies are uniformly distributed on the sky. (See Fig. \ref{fig:bias} for an illustration). 

To overcome the above selection bias in photometric CG samples, \citet{1996AJ....112..871B} introduced an alternative approach: the friend-of-friend (FoF) algorithm, which identifies whether two galaxies belong to the same group using their projected separation and line-of-sight (LOS) velocity difference. Several authors identified CGs following this method from various redshift surveys, including CfA2 \citep{1996AJ....112..871B} and 2dFGRS \citep{2007ggnu.conf...73S}, and from the SDSS main galaxy sample (MGS) \citep{2016ApJS..225...23S}. This FoF algorithm has a strong dependence on the redshift completeness of the sample galaxies. Modern galaxy redshift surveys typically use a multifiber spectrograph, where the fiber collision effect brings an incompleteness effect at small angular scale. For example, the fibers cannot be placed closer than 55" on the same plate in the SDSS \citep{2002ApJ...571..172Z}, which leads to a very high spectroscopic incompleteness on the compact galaxy system (\citealp{2008ApJ...685..235P,2016RAA....16...43S}). Moreover, distant CGs tend to be smaller in angular sizes and therefore are more likely to be biased by  this incompleteness effect.

Recently, \citet{2018AA...618A.157D} combined the advantages of the above two selection algorithms and applied their new algorithm on the galaxy catalog of SDSS-DR12 \citep{2015ApJS..219...12A}. Specifically, they first apply a redshift filter to remove the background galaxies. These galaxies without redshifts are considered as background at this step. They then select the CG candidates inside the redshift slices using the Hickson's criteria. Finally, they bring those galaxies without redshifts back and check whether they could be the interlopers of these spectroscopic CG candidates. These groups without interlopers are finally confirmed as  `noncontaminated' CGs and otherwise listed as `potentially contaminated' CGs. With the redshift slice to remove the background galaxies in the beginning, this algorithm effectively avoids the selection bias caused by the low redshift CGs in the traditional photometric-only selection technique.    

However, in the algorithm of \citet{2018AA...618A.157D}, all the CG samples are initially selected based on spectroscopic galaxy samples. A real CG with spectroscopic incompleteness will be missed in the final CG sample if its spectroscopic members cannot pass the Hickson's criteria. To overcome this, we propose a revised algorithm to select CGs, where we keep rather than remove these galaxies without redshift measurements during the initial CG selection (see Section \ref{sec:selection} for detail). With this new algorithm,  we tend to retain all CG candidates, which could be easily verified with future spectroscopic survey, e.g. the complementary galaxies in the LAMOST spectral survey \citep{2004ChJAA...4....1S,2012RAA....12.1197C,2012RAA....12.1243L}.  Actually, the redshift completeness of the SDSS main sample galaxies have already been significantly improved by the LAMOST spectral survey \citep{2015RAA....15.1095L,2016RAA....16...43S,2019ApJ...880..114F}. 

Therefore, the motivation of this study is to take the advantages of the new selection algorithm and the supplied redshifts from the LAMOST spectral survey to build the most complete and well-defined low-$z$ CG samples. In the next paper of this series, we will use this CG sample to study their dynamical properties and  environmental dependence  in detail. This paper is organized as follows. In section~\ref{sec:mgs}, we briefly describe the galaxy catalog used in this work. In section~\ref{sec:selection}, we optimize the selection procedure for maximizing the CG samples and preserving all possible CG candidates. We describe the derived CG catalogs in section~\ref{sec:CG-Cats} and make comparisons with other available CG catalogs in section~\ref{sec:compare}. We provide a brief discussion on the final CG samples in section~\ref{sec:discussion} and finally summarize our results in section~\ref{sec:summary}. Throughout this paper, we assume the flat WMAP7 cosmology with parameters: $H_{0} = 70$ km s$^{-1}$ Mpc$^{-1}$ and $\Omega_{m} = 0.27$ \citep{2011ApJS..192...18K}.

\begin{deluxetable}{lr}
\tablecaption{Sources of Redshift Measurement \label{tab:extraZ}}
\tablewidth{\columnwidth}
\tablehead{
\colhead{Redshift Survey \tablenotemark{}} &
\colhead{Redshift Number \tablenotemark{}}} 
\startdata
SDSS-DR14 & 694,930 \\
LAMOST (Till DR7-Q2) & 8,021 \\
GAMA-DR2 & 1,017 \\
Others from VAGC & 8,231 \\
\hline
Total Redshift Measurement & 712,199 \\
Total Galaxy Sample & 746,950 
\enddata
\end{deluxetable}

\section{Galaxy Samples} \label{sec:mgs} 
In this paper, we select the CGs in the largest and most complete low redshift spectroscopic galaxy sample, the MGS in the legacy of the SDSS, which is defined as the galaxies with $r$-band Galactic extinction corrected Petrosian magnitude  $r \leq 17.77$ \citep{2002AJ....124.1810S}. In SDSS, the bright galaxies are not complete in either photometric (deblending effect) or spectroscopic (saturation effect) sample \citep{2002AJ....124.1810S}. Therefore, we also impose a bright-end limit $r \ge 14.00$ on our sample galaxies, following \citet{2004AJ....127.1811L}. We take the basic photometric parameters and spectroscopic redshifts of the MGS from the NYU Value-Added Galaxy Catalog \citep[][hereafter NYU-VAGC]{2005AJ....129.2562B}, which is based on DR7 of the SDSS \citep{2009ApJS..182..543A}. 

Until the SDSS-DR7, there were about $\sim 7.0 \%$  of the MGS lacking spectroscopic redshifts due to the fiber collision effect. In NYU-VAGC, besides the spectroscopic redshifts from the SDSS DR7, extra redshifts are collected  from 2dfGRs \citep{2001MNRAS.328.1039C}, 2MASS \citep{2006AJ....131.1163S}, PSCz \citep{2000MNRAS.317...55S} and RC3 \citep{1991rc3..book.....D}. After DR7, the galaxies in the MGS without spectroscopy are continually targeted in the SDSS, which are also are targeted as a complementary galaxy sample in the LAMOST spectral survey \citep{2015RAA....15.1095L}.  Following \citet{2019ApJ...880..114F}, we matched the photometric MGS sample with the SDSS-DR14, the most up-to-date LAMOST data release (DR7 V0, until March of 2019) and the GAMA-DR2 \citep{2016yCat..74522087L} and obtained a significant amount of extra redshifts. For these galaxies with more than one spectroscopy redshift, we set the priority as follow: SDSS $>$ LAMOST $>$ GAMA $>$ others from VAGC. Basically, the galaxy sample used in this study is an updated version of \citet{2019ApJ...880..114F} (updated with the newest data release of the LAMOST spectral survey), which contains $746,950$ galaxies and has a spectroscopic completeness of $\sim 95.3\%$. The detailed numbers of the global galaxy catalog and their spectroscopic redshifts from different surveys are listed in Table~\ref{tab:extraZ}.

\section{Compact Groups} \label{sec:selection}
\subsection{Selection Criteria} \label{sec:Criteria}
In our CG sample selection, we slightly revise Hickson's criteria in the following way:

\begin{enumerate}
\setlength{\itemsep}{-0.7ex}
\item Richness: $3\leq N$ $(14.00 \leq r \leq 17.77)$ $\leq 10$
\item Isolation: $ \theta_{n} \geq 3$ $\theta_{G}$ 
\item Compactness: $ \mu \leq 26.0$ mag\,arcsec$^{-2}$
\item Velocity Difference: $|V - V_{\text{med}}| \leq 1000$ km s$^{-1}$ 
\end{enumerate}

where $V$ is the radial velocity of each member galaxy and $V_{\text{med}}$ is the median radial velocity of the members. Our new selection criteria are different from the traditional Hickson's one on two aspects. First, considering the fact that the triplet system is not distinguished from $N \geq 4$ groups \citep{2013MNRAS.433.3547D}, we extend the first criterion so as to include the triplets and maximize the sample size. We add an additional upper limit of $N \leq 10$ to distinguish the groups from rich clusters and accelerate the running time of the searching algorithm. Since the maximum richness of our derived groups contain $N = 8$ member galaxies, this is not a very strict constraint. Second, we remove the constraint on the magnitude range $\Delta m \leq 3$, which is based on the consideration that the galaxies with similar magnitudes are more likely to be at similar redshifts. Here, our CG selection is based on spectroscopic galaxies, so the magnitude constraint is no longer necessary. Moreover, since our source galaxy catalog is in the magnitude range $14.00 \leq r \leq 17.77$, the resulting CGs will have $\Delta m_{r,max} = 3.77$ and therefore they are quite comparable to the traditional Hickson's CGs (see more discussion on this criteria in Section \ref{sec:discussion}).

\subsection{Selection Procedure} \label{sec:procedure}

\begin{figure}
\begin{center}
\includegraphics[width=\columnwidth]{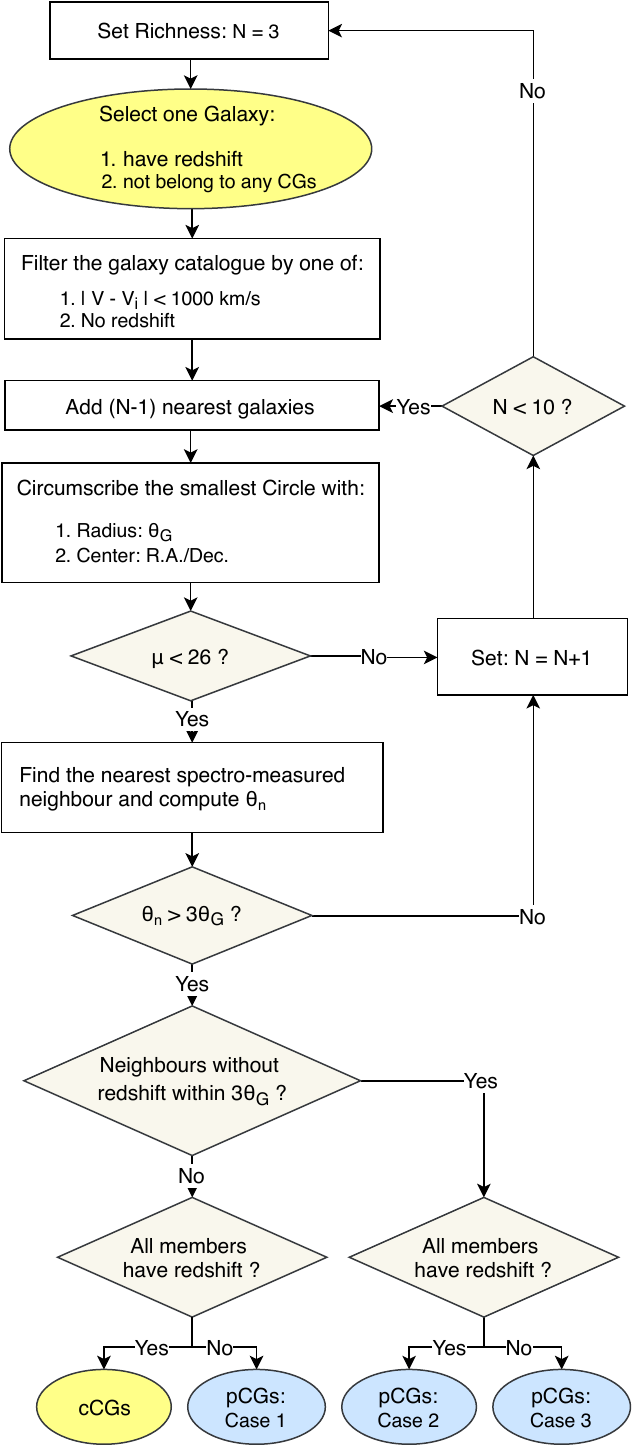}
\caption{Flow chart of our CG selection procedure. \label{fig:flowchart}}
\end{center}
\end{figure}

\begin{figure}
\begin{center}
\includegraphics[width=\columnwidth]{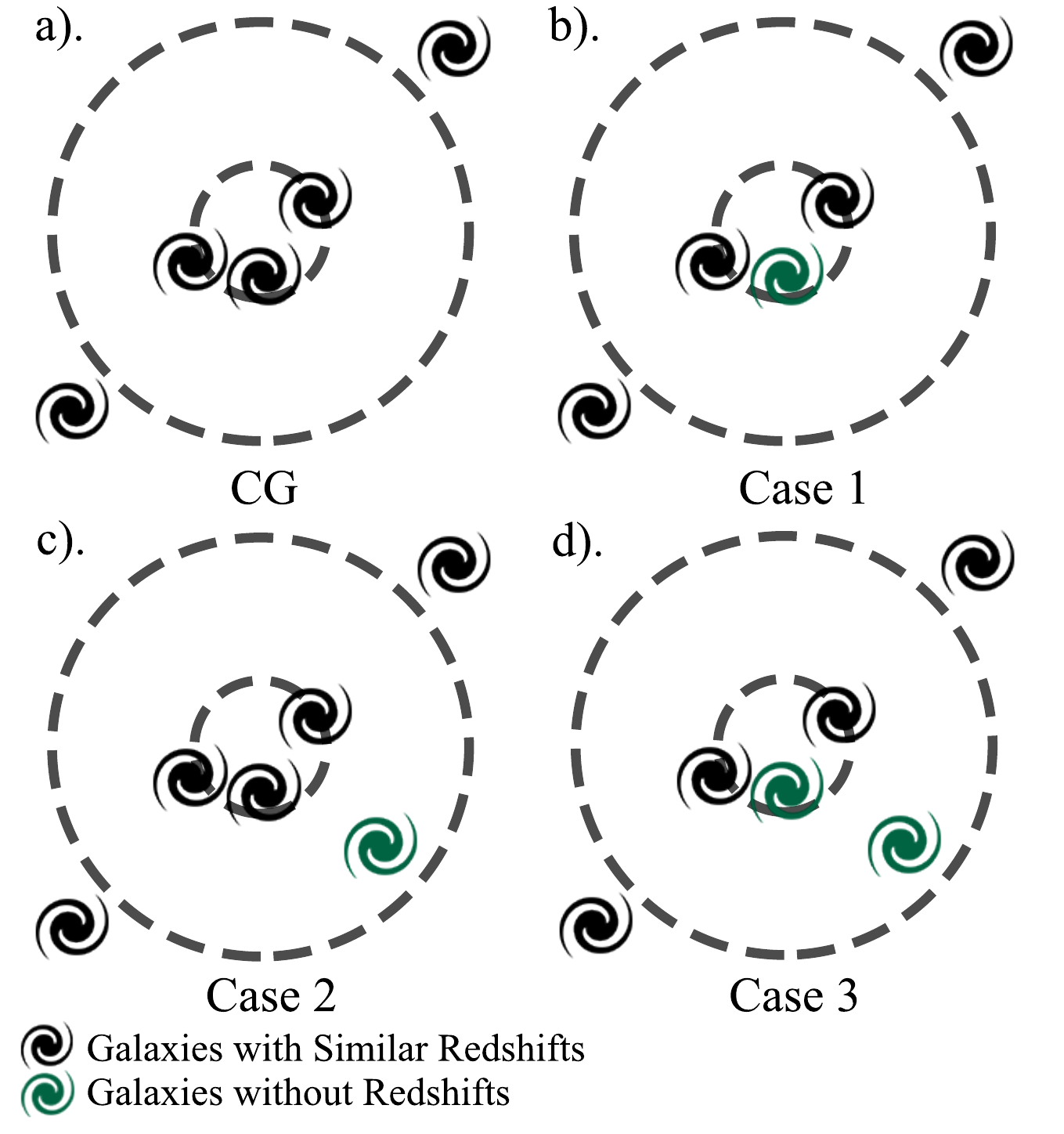}
\caption{Cartoons of the various cases of the CGs and CG candidates. (a) Conservative sample of compact groups: meet all requirement of CGs: meets all requirements of CG criteria. (b) Case 1: meets all CG criteria except that at least one member galaxy lacks spectroscopic redshifts. (c) Case 2: meets all CG criteria redshifts except that there are galaxies without spectroscopic redshift inside the isolation ring. (d) Case 3: Combination of Cases 1 and 2.\label{fig:case}}
\end{center}
\end{figure}

\begin{deluxetable}{lccr}
\tablecaption{The total numbers of the cCGs and pCGs and their member galaxies. \label{tab:lists}}
\tablecolumns{4}
\tablewidth{\columnwidth}
\tablehead{
\multicolumn{2}{c}{Catalog} &
\colhead{Groups\tablenotemark{}} &
\colhead{Member Galaxies\tablenotemark{}}}
\startdata
cCGs & ... & 6,144 & 19,465 \\
\hline
& Case 1 & 6,790 & 21,382 \\
pCGs & Case 2 & 458 & 1,389 \\
& Case 3 & 774 & 2,361 \\
\cline{2-4}
& Total & 8,022 & 25,132 
\enddata
\end{deluxetable}

We show the flow chart of the selection procedure of the CGs and CG candidates in Fig.\ref{fig:flowchart}. The main steps are outlined as follows:

{\textcolor{gunen}{Step 0}}: Start from a galaxy with spectroscopic redshift (recessional velocity $V_i$) in turn.

{\textcolor{gunen}{Step 1}}: Filter out the foreground and background galaxies and only keep galaxies with $|V-V_i| < 1000 \rm{km/s}$. In this step, these galaxies in the same magnitude range but without redshift measurements are considered as the same recessional velocity as $V_i$.

{\textcolor{gunen}{Step 2}}: Find the smallest circle that encircles $3\le N \le 10$ members. Write the radius of this circle as $\theta_G$ and mark the center of this circle as the coordinate of the candidate CG. Calculate the mean surface brightness inside the circle and check the compactness criterion. 

{\textcolor{gunen}{Step 3}}: Find the nearest neighbor galaxy with spectroscopic redshift, write the distance to the CG center as $\theta_n$. In this step, all the galaxies in the neighboring region without redshifts are considered as background galaxies. Check the isolation criteria. These CG candidates do not pass the isolation criteria, set $N = N+1$ and go back to Step 2. 

{\textcolor{gunen}{Step 4}}: Separate CG candidates into 4 different catalogs according to the redshift status of galaxies. One is the conservative sample of CGs (hereafter cCGs) with spectroscopic redshifts for all the members and the nearest neighbor, the others with assumed redshifts are named as possibly CGs (hereafter pCGs). For the pCGs, there are three different kinds of cases.

\begin{enumerate}
\setlength{\itemsep}{-0.7ex}
\item {\bfseries Case 1}: At least one of the members lacks redshift data but the others have genuine spectroscopic redshifts.
\item {\bfseries Case 2}: All the members have genuine spectroscopic redshifts but at least one galaxy without redshift lies in the isolation rings.
\item {\bfseries Case 3}: The combination of Cases 1 and 2.
\end{enumerate}

The cartoon of the configurations of four types of CGs are shown in Fig.~\ref{fig:case}. In our selection procedure, we do not consider the extreme case, where one CG is possibly embedded in a larger CG, i.e., our selection procedure stops once a CG or CG candidate is identified.

{\textcolor{gunen}{Step 5}}: Visual inspections on the images of all CGs.  With this step, we remove some fake sources from bad photometry \citep[e.g.][]{2009MNRAS.395..255M}. Most of the contamination cases are caused by the effect that a large extended galaxy is split into many smaller parts and identified as isolated galaxies respectively.

With the above procedure, we finally obtain 6,144 cCGs and 8,022 pCGs. The detailed number of each type CGs and the number of their members are listed in Table~\ref{tab:lists}.

For illustration, we show example SDSS images of each type CG and CG candidate in Appendix~\ref{sec:example}.

\subsection{Velocity Dispersion of cCGs} \label{sec:veldisper}
We estimate the rest-frame LOS velocity dispersion of our samples of CGs using the gapper estimator \citep{1990AJ....100...32B}. \citet{2012MNRAS.426..296D} suggests that the gapper estimator shows much fewer biases for groups with small numbers of members than the standard estimator of dispersion. For the ordered set of recession velocities $\{V_i\}$ of $N$ group members, the gaps are defined by:

\begin{center}
$g_{i} = V_{i+1} - V_{i}, \qquad i = 1, 2, \cdots, N $ \,,
\end{center}

and the rest-frame velocity dispersion is estimated by
$$ \sigma_{\text{LOS}} = \frac{\sqrt{\pi}}{\left(1 + z_{g}\right) N(N-1)} \sum_{i=1}^{N-1} w_{i}g_{i} $$ \,,
where $z_{g}$ is the group redshift and $w_{i}$ is the Gaussian weight defined as $w_{i} = i(N-i)$. 

\subsection{Compact Group Catalogs} \label{sec:CG-Cats}
We present the final catalogs of 6,144 cCGs and 8,022 pCGs in Table ~\ref{tab:cCGs} and ~\ref{tab:pCGs}, respectively.

In these two tables, we list their Group ID, sky coordinates (R.A. \& decl.), redshift, richness, angular radius, surface brightness and $\Delta m \le 3$ criterion flag for each table. For each CG, the sky coordinate and angular radius are the center and radius of the minimum circle that encircles the group members. The surface brightness is thus the average surface brightness of the group members inside that circle. The redshift is the average redshifts of all group members, where the members in the pCGs with our redshift measurements have not been taken into account. The $\Delta m \le 3$ criterion flag indicates whether all the group members are inside a magnitude range $\Delta m \le 3$ (`0' for `False'; `1' for `True', see more discussion in section~\ref{sec:discuss-deltav}). In table~\ref{tab:cCGs}, we list the LOS velocity dispersion of each CG, while in table~\ref{tab:pCGs} we provide the `case flag' to show which groups belong to which case of pCGs as that demonstrated in Fig.~\ref{fig:case}. 

Table~\ref{tab:cCGs-m} and table~\ref{tab:pCGs-m} list the properties of member galaxies of each cCGs and pCGs, respectively, including group ID, member ID, sky coordinates, redshift, Galactic extinction corrected $r$-band Petrosian magnitude, and spectroscopy data source. In table~\ref{tab:pCGs-m}, we also list the possible interlopers inside the isolation ring (i.e. $\theta_G < \theta < 3 \theta_G$) of Case 2 and Case 3 pCGs, which have been assumed as background galaxies during the CG identification for completeness. These galaxies are assigned with member ID `$-99$' for its corresponding groups. With this information, the Case 2 and Case 3 pCGs could be easily identified with future spectroscopic redshifts. For each catalog, only a few part of them are listed here. Full versions of these tables are available online in readable format.

\begin{figure*}
\begin{center}
	\includegraphics[width=\textwidth,height=\textheight,keepaspectratio]{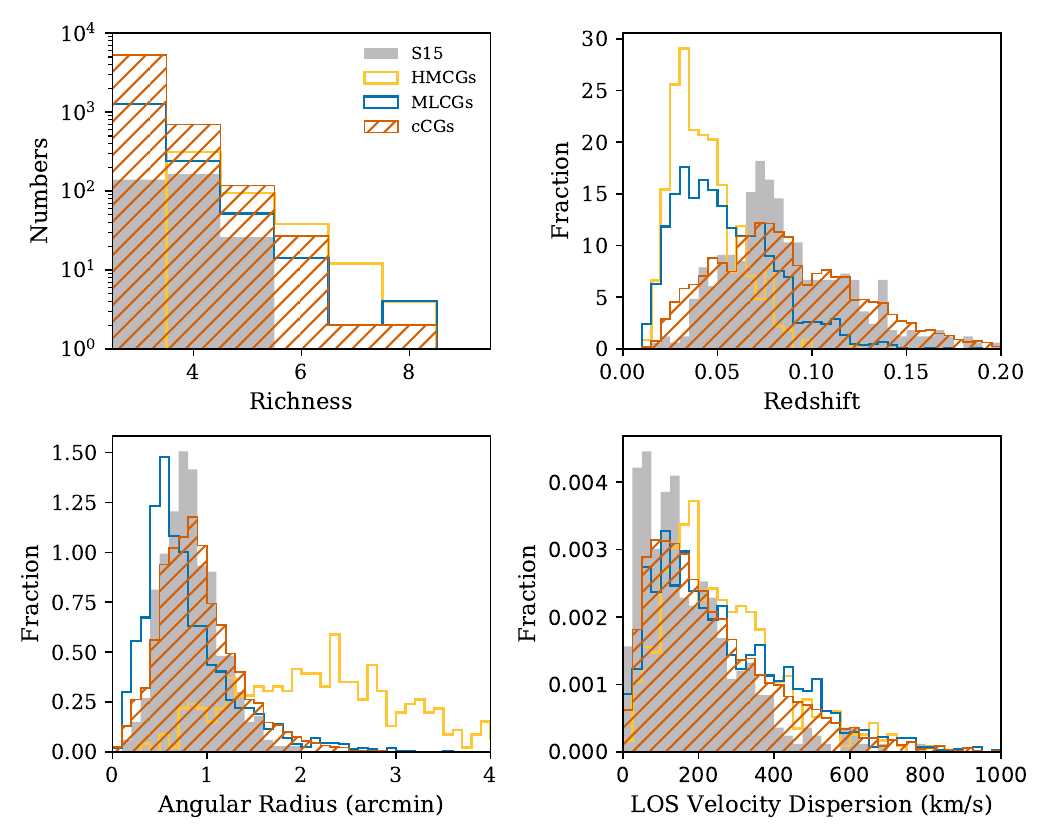}
    \caption{Comparison of the distributions of the CGs in various catalogs: cCGs (red hatched), HMCG (blue), MLCG (yellow), and S15-A (gray shaded). Upper left: group richness. Upper right: group redshift. Lower left: angular radius of the smallest enclosed circle. Lower right: velocity dispersion of the group. \label{fig:Props}}
\end{center}
\end{figure*}

We show the basic statistical properties of our cCG catalog in Fig.\ref{fig:Props}, where the distributions of the richness ($N$), redshift ($z_{G}$), angular radius ($\theta_{G}$) and velocity dispersion ($\sigma_{\text{LOS}}$) are plotted as the hatched histograms in the upper left, upper right , lower left, and lower right panel respectively. The richness of our cCGs spans a range from 3 to 8 and their redshift distribution peaks at $z\sim 0.08$. The typical angular radius of the cCGs is about $\sim 1^{\prime}$, indicating their compact nature. The distributions of these three apparent parameters are primarily resulted from the magnitude limit of our galaxy sample ($14.00 \leq r \leq 17.77$) and our CG selection algorithm. We will present a detailed comparison of our cCGs with other CG catalogues that selected from the same SDSS galaxy sample on these distributions in Section~\ref{sec:compare}. The velocity dispersion $\sigma_{\text{LOS}}$, which characterizes the dynamical properties of the galaxy groups, is more robust against the selection effects. Our cCGs show similar $\sigma_{\text{LOS}}$ distribution as other catalogues in the range from 50 to 600 km/s, which is basically consistent with the expectation of normal galaxy groups (see more discussions in Section~\ref{sec:discuss-deltav}).

\begin{deluxetable*}{lcccccccc}
\tablecaption{Catalog of the cCGs, which includes the Columns of Group ID, sky coordinates, richness, redshift, angular radius, surface brightness, LOS velocity dispersion, and $\Delta {m} \le 3$ criterion flag (`0' for `False'; `1' for 'True').  \tablenotemark{} \label{tab:cCGs}}
\tablewidth{2\columnwidth}
\tablehead{
\colhead{Group ID} & \colhead{R.A.} & \colhead{Decl.} & \colhead{N} & \colhead{$z_{g}$} & \colhead{$\theta_{G}$} & \colhead{$\mu$} & \colhead{$\sigma_{\text{LOS}}$} & \colhead{{\bfseries $\Delta {m} \le 3$}\tablenotemark{}} \\
\colhead{} & \colhead{(J2000)} & \colhead{(J2000)} & \colhead{} & \colhead{} & \colhead{(arcmin)} & \colhead{(mag/arcsec$^{2}$)} & \colhead{(km/s)} & \colhead{}
}
%\colnumbers
\startdata
cCGs-0001 & 56.1382 & 1.0558 & 3 & 0.10425 & 0.526 & 24.265 & 484.706 & 1\\
cCGs-0002 & 57.2751 & 0.8998 & 3 & 0.10925 & 0.993 & 25.965 & 68.366 & 1\\
cCGs-0003 & 239.8562 & -0.9450 & 3 & 0.10238 & 0.463 & 24.133 & 340.199 & 1\\
cCGs-0004 & 169.7778 & -0.2108 & 3 & 0.09746 & 0.466 & 24.185 & 398.755 & 1\\
cCGs-0005 & 241.2247 & -0.0601 & 3 & 0.05093 & 1.328 & 25.784 & 287.055 & 1\\
cCGs-0006 & 246.1077 & 0.8071 & 3 & 0.05834 & 1.160 & 25.681 & 503.839 & 1\\
cCGs-0007 & 174.9150 & -1.0894 & 3 & 0.07777 & 0.622 & 24.721 & 412.498 & 1\\
cCGs-0008 & 176.5369 & -1.1010 & 5 & 0.11765 & 1.200 & 25.475 & 390.835 & 1\\
cCGs-0009 & 177.1710 & -1.1997 & 3 & 0.10660 & 0.312 & 23.896 & 415.169 & 1\\
cCGs-0010 & 177.2054 & -1.1776 & 3 & 0.10647 & 0.488 & 24.590 & 167.878 & 1
\enddata
\tablenotetext{}{\textcolor{gunen}{Note.} This table has 6144 rows, of which only the first 10 rows are displayed here.}
\end{deluxetable*}

\begin{deluxetable*}{lcccccccc}
\tablecaption{Catalog of the pCGs, which includes the Columns of Group ID, sky coordinates, richness, redshift, angular radius, surface brightness, the classification of uncertain case as described in Fig.~\ref{fig:case}, and $\Delta {m} \le 3$ criterion flag. \tablenotemark{} \label{tab:pCGs}}
\tablewidth{2\columnwidth}
\tablehead{
\colhead{Group ID} & \colhead{R.A.} & \colhead{Decl.} & \colhead{N} & \colhead{$z_{g}$} & \colhead{$\theta_{G}$} & \colhead{$\mu$} & \colhead{Case} & \colhead{{\bfseries $\Delta {m} \le 3$}\tablenotemark{}}\\
\colhead{} & \colhead{(J2000)} & \colhead{(J2000)} & \colhead{} & \colhead{} & \colhead{(arcmin)} & \colhead{(mag/arcsec$^{2}$)} & \colhead{} & \colhead{}
}
\startdata
pCGs-0001 & 241.3722 & -0.5285 & 3 & 0.13029 & 0.555 & 24.924 & 1 & 1\\
pCGs-0002 & 242.1523 & -0.1234 & 3 & 0.13994 & 0.888 & 25.946 & 2 & 1\\
pCGs-0003 & 242.6850 & -0.1302 & 3 & 0.10632 & 0.846 & 25.403 & 1 & 1\\
pCGs-0004 & 239.8296 & 0.3824 & 4 & 0.13990 & 1.322 & 25.898 & 1 & 1\\
pCGs-0005 & 239.8340 & 0.3660 & 4 & 0.09341 & 1.123 & 25.410 & 1 & 1\\
pCGs-0006 & 242.6550 & 0.3033 & 3 & 0.08022 & 0.568 & 24.665 & 3 & 1\\
pCGs-0007 & 242.6661 & 0.3009 & 3 & 0.05806 & 0.367 & 23.098 & 1 & 1\\
pCGs-0008 & 242.7184 & 0.3139 & 3 & 0.10622 & 0.736 & 24.834 & 1 & 1\\
pCGs-0009 & 242.5358 & 0.7680 & 3 & 0.04192 & 0.204 & 20.544 & 1 & 1\\
pCGs-0010 & 243.3732 & 0.8159 & 3 & 0.08196 & 0.419 & 24.480 & 1 & 1
\enddata
\tablenotetext{}{\textcolor{gunen}{Note.} This table has 8022 rows, where only the first 10 rows are displayed.}
\end{deluxetable*}

\section{Comparison with other CG Catalogs} \label{sec:compare}
In this section, we compare our cCGs with other CG catalogs derived from the SDSS.

\begin{enumerate}
\setlength{\itemsep}{-0.7ex}
\item \textcolor{gunen}{S15}: \citet{2009MNRAS.395..255M} derived the CG samples using Hickson criteria from the SDSS-DR6 photometric catalog \citep{2008ApJS..175..297A} in two ranges of magnitude limit and resulted in two data sets, Catalog A and B. Later, \citet{2015JKAS...48..381S} supplied redshifts from FLOW/FAST observations and the SDSS-DR12, using the velocity filter to check the candidates of Catalog A. The final filtered sample (S15, hereafter) is comprised of 332 CGs with at least three member galaxies. This catalog is an implementation of the traditional searching procedure of CGs.
\item \textcolor{gunen}{MLCG}: \citet{2016ApJS..225...23S} applied an FoF algorithm on the enhanced SDSS-DR12 magnitude-limited redshift survey and extracted a catalog of 1,588 CGs (MLCG, hereafter). Their FoF algorithm found the neighboring galaxies within a fixed projected distance and velocity difference. They also applied the similar richness and compactness criteria as used by \citet{2008MNRAS.387.1281M}, but the isolation criterion was not taken into account.
\item \textcolor{gunen}{HMCG}: As we already mentioned, \citet{2018AA...618A.157D} applied a redshift filter on galaxy sample before starting the Hickson's criteria to search CGs, where a sample of 462 CGs are finally extracted from the SDSS-DR12.
\end{enumerate}

We summarize the methodology of the selection algorithm and the main features of the above CG catalogs in table~\ref{tab:summerize}.  We show the distributions of the richness, redshift, angular radius, and LOS velocity dispersion of the CGs in these catalogs in Fig.~\ref{fig:Props}. Next, we discuss and compare them with our cCGs one by one.

\subsection{cCGs versus S15} \label{sec:S15-A}
The identification of S15 is consistent with the implementation of the traditional Hickson's selection procedure. As expected, the relative fraction of nearby groups is lower than cCGs as shown in the upper right panel of Fig.\ref{fig:Props}.

To make a detailed comparison with S15, we remove 52 CGs from S15  sample, which have at least one member galaxy without spectroscopic redshifts and would be identified as Case 1 pCGs in our study. We match the remaining 280 S15 CGs with  our cCGs based on the angular separation and radial velocity difference and find 220 out of 280 ($\sim$ 78.6\%) overlaps. The mismatched groups are mainly attributed to the small differences in the selection criteria, where S15 draws the galaxy catalog with $14.50 \leq r \leq 18.00$ and takes the restrictive condition $\Delta m \leq 3$ of the standard Hickson's criteria (see more discussions in Section \ref{sec:discussion}).

\subsection{cCGs versus MLCGs} \label{sec:S16}
For MLCGs, as we have introduced, the FoF selection algorithm biases against the high redshift objects in the SDSS.  The projected distance restriction used in MLCGs is $D_{\text{lim}} = 50 h^{-1} $ kpc, which corresponds to $55^{\prime}$ at redshift $z \sim 0.07$. Also, the exclusion of the isolation criterion and the inclusion of very bright galaxies even make the redshift distribution of the MLCGs  be biased toward low redshifts as the  upper right panel of Fig.\ref{fig:Props} shows.

We also match the MLCGs with our cCGs and find that 736 out of 1,588 ($\sim$ 46.3\%)  MLCGs overlap with cCGs. This low match rate is mainly attribute to the neglect of isolation criterion in MLCGs. As mentioned in \citet{2016ApJS..225...23S}, 1,228 ($\sim$ 77.0\%) MLCG systems violated Hickson's original isolation criterion.  We find that 352 MLCGs violate our modified isolation criterion. Another cause of the difference is the inclusion of the  bright ($r \le 14.00$) galaxies in MLCGs, which have been excluded in our study. There are 239 ($\sim$ 15.1\%) MLCGs that contain these very bright members, most of them are located very nearby ($z \le 0.05$).  That is to say, MLCGs are a very good complementary sample to our CGs, especially at low redshifts.

\subsection{cCGs versus HMCGs} \label{sec:D18}
HMCGs comprises 406 noncontaminated CGs and 56 potentially contaminated CGs. Although our study shares similar selection algorithm and galaxy catalog as those of HMCGs, the differences between two samples are significant. HMCGs are selected with a strict magnitude limit criterion that the brightest CG member should be at least three magnitudes brighter than the completeness magnitude of the galaxy sample ($r_b \leq r_{\text{lim}}-3$), which ensures the homogeneity of their CGs at different redshift (see further discussion in Section~\ref{sec:discuss-ccgs}). 
Considering $r_{\text{lim}} \sim 17.77$ for the SDSS MGS, the brightest galaxy members of all HMCGs are therefore brighter than 14.77. Moreover, during the selection of HMGCs, there is no bright magnitude limit on the sample galaxies, while our CGs are selected on galaxies with $r \geq 14.00$. Combining these two effects, the redshift distribution of HMCGs is significantly biased to lower redshifts as the upper right panel of Fig.\ref{fig:Props} shows. The third difference of the HMCG selection from our study is that they adopt a fainter brightness criterion down to $ \mu \leq 26.33$ mag\,arcsec$^{-2}$ . Combining the low redshift selection bias and the lower surface brightness criterion, HMCGs have significantly larger angular diameters than all other CG catalogs (lower left panel of Fig.\ref{fig:Props}).

To make a fair comparison, we remove the HMCGs that either contain very bright members ($r < 14.00$) or with the surface brightness $ \mu \geq 26.00$ mag\,arcsec$^{-2}$ and  218 groups are remain. We cross-match these 218 HMCGs with our cCGs and find 121 of them are overlapped. The other HMCGs that not listed in our cCGs are mainly  caused by the wider  magnitude range of the sample galaxies ($14.00 \leq r \leq 17.77$) used in our CG selection than in HMCGs ($r_b \leq r \leq r_b+3$).

These additional galaxies ($r_b+3 \le r \leq 17.77$), depending on their redshifts,  might change the result of CG selection. If these additional galaxies were background galaxies, that would not change the CG identification. On the other hand, if these additional galaxies had accordance redshifts with the corresponding CGs, then they would be absorbed as the CG members and might not pass the isolation criterion  in the next step of our CG selection. We show an example of such a case in Figure \ref{fig:range3} of the Appendix \ref{sec:example} for illustration.  

The differences among the final CG samples,  resulted from the subtle differences of the selection criteria, indicates that the selection of a unique CG sample is quite nontrivial.  

\begin{figure}
\begin{center}
	\includegraphics[width=\columnwidth,keepaspectratio]{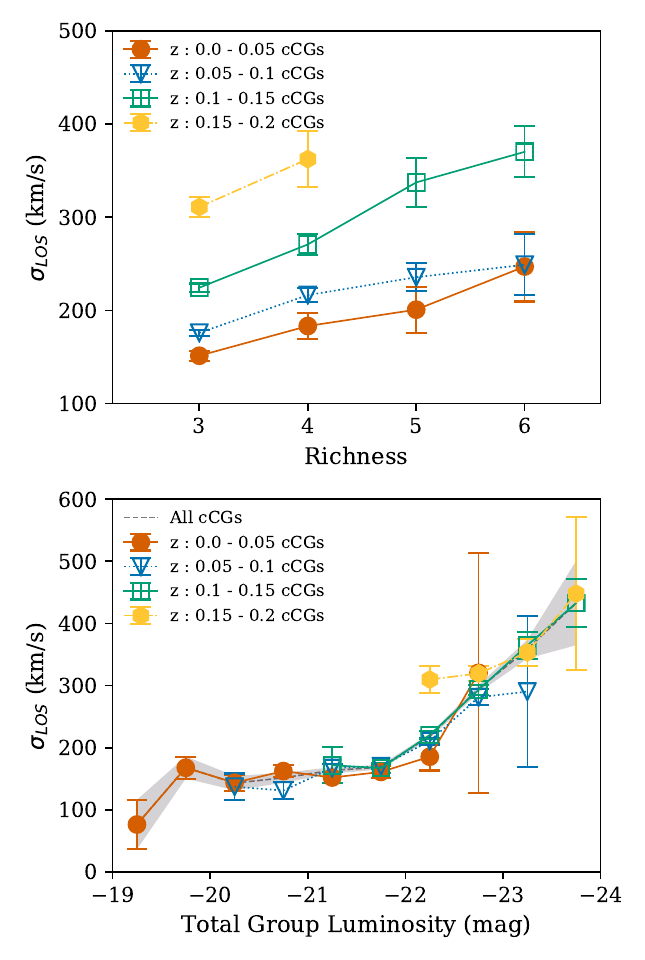}
    \caption{Median LOS velocity dispersion ($\sigma_{\text{LOS}}$) and its uncertainty as a function of group richness (upper panel) and total luminosity of observed members (lower panel) for cCGs at different redshift bins. Different symbols represent the CGs in different redshift bins as labeled in the top left corner of the figure. In the lower panel, the gray shaded area represents the same relation of all cCGs. \label{fig:flim}}
\end{center}
\end{figure}

\section{Discussion} \label{sec:discussion}
\subsection{Selection Criteria} \label{sec:discuss-deltav}
As we have shown in the bottom right panel of Fig.~\ref{fig:Props}, the peak of the $\sigma_{\text{LOS}}$ distribution of our cCGs is smaller than 200 km s$^{-1}$. In our CG selection, we use a velocity difference criteria $\Delta V = |V - V_{\text{med}}|$ smaller than 1000 km s$^{-1}$. In general, this radial velocity difference restriction minimizes the interlopers with discordant redshifts while recovering systems similar to the original Hickson CGs \citep{2010AJ....139.1857W}. However, this large critical $\Delta V$ value brings doubt that it might be too large for low mass groups (e.g. these with $\sigma_{\text{LOS}} < 200$ km s$^{-1}$). To test this effect, for each cCG, we compare $\Delta V$ of each group member with its $\sigma_{\text{LOS}}$ and find most of our groups have members with $\Delta V < 2 \sigma_{LOS}$ and none of them with $\Delta V > 3 \sigma_{\text{LOS}}$. This result may not be surprising. On the one hand, for CGs with few members, any group member with a large $\Delta V$ value would also bias $\sigma_{LOS}$ to a large value. On the other hand, as we will show in the next section, $\sigma_{\text{LOS}}$ is nicely correlated with the total luminosity of the group members, which indicates that $\sigma_{\text{LOS}}$ is a good dynamics indicator and therefore our cCGs could not be significantly contaminated by possible interlopers. 

Nevertheless, we emphasize the critical $\Delta V$ value 1000 km s$^{-1}$ we adopt is somewhat arbitrary. A minor revision of this critical value will also slightly change our final CG catalog. For example, if we  apply a tighter critical value, $\Delta V < 800$ km s$^{-1}$,  about $\sim 5 \%$ groups  would be removed from current cCG sample. The critical value 1000 km s$^{-1}$ we take is for the consistency with other studies, and also makes the comparison in Section~\ref{sec:compare} easier.

Also, during the construction of the CG samples, we do not restrict the group members within a magnitude range $\Delta m \le 3$ as that typically used in other studies (e.g. \citealp{2009MNRAS.395..255M}). As we mentioned in Section~\ref{sec:Criteria}, the traditional Hickson criterion $\Delta m \le 3$ is applicable to photometric-only galaxy samples, where the galaxies with similar magnitudes are more likely to have accordant redshifts. While our study is based  on  spectroscopic galaxy sample, we therefore do not need $\Delta m \le 3$ to ensure the group nature of the selected galaxies.
In our study, all CGs are selected in the magnitude range $14.00 \leq r \leq 17.77$ so that all of them have $\Delta m \le 3.77$ and are comparable to the $\Delta m \le 3$ that used in other studies (e.g. S15, HMCGs). Also, most of our CGs are located at $z > 0.03$ and their brightest members have $r_b \geq 14.77$, which in turn makes most of the cCGs also satisfy $\Delta m \le 3$. In fact, only 196 cCGs ($\sim 3.2 \%$) and 297 pCGs ($\sim 3.7 \%$) violate the $\Delta m \le 3$ criterion. Apparently, these groups with $\Delta m \geq 3$ might be different from the traditional CGs. We therefore add a flag in the last column of Table~\ref{tab:cCGs} and ~\ref{tab:pCGs} to call attention. We keep these groups with $\Delta m > 3$ in our CG catalog based on two considerations. First, our CGs are selected from a magnitude-limited sample and the keeping of all the members inside the magnitude range $14.00 \leq r \leq 17.77$ makes the correction of selection effect (our next study) easier. Second, the compact nature of the CGs is ensured by the mean surface brightness of the group system (the compactness criterion), which is weakly affected by the inclusion of the faint galaxies. That is to say, the increasing of $\Delta m \le 3$ to $\Delta m \le 3.77$ has little impact on the compact nature of the selected groups.

Finally, we emphasize, for the magnitude-limited galaxy sample, a simple $\Delta m$ criterion (no matter its specific value)  brings a significant inhomogeneity effect for groups at different redshifts. We make a detailed discussion next.

\subsection{Inhomogeneity of cCGs} \label{sec:discuss-ccgs}
Our CGs are derived from a magnitude-limited sample of galaxies by applying a modified algorithm of the traditional Hickson Criteria, which brings a redshift dependent bias that makes the magnitude range of the group members be a function of its redshift. In fact, such a bias exists in all the CG samples based on magnitude-limited galaxy sample (e.g., S15, MLCG) except HMCG, where the brightest group galaxies are further required to fulfill $r_b \le r_{\text{lim}} - 3$. However, such a strong restriction in HMCG also makes the sample size small. Our study aims to maximize the CG sample size yet with `well-defined' selection criteria and leave the correction of the sample selection effect in the upcoming study. Related to this, as we have discussed in Section~\ref{sec:Criteria}, the relaxing of  the traditional $\Delta m<3$ criterion to $14.00 \le r \le 17.77$ is also motivated by this consideration.

The inhomogeneity effect also makes the richness being a biased indicator of the group mass at different redshifts, as the upper panel of figure~\ref{fig:flim} shows; the median LOS velocity dispersion of cCGs ($\sigma_{\text{LOS}}$) is plotted as a function of their richness in different redshift bins. As can be seen, at a given richness, the median LOS velocity dispersion is systematically higher at higher redshifts. 

To alleviate this inhomogeneity effect, it is better to use the total group luminosity rather than the richness to characterize the global group mass. A detailed calculation of the total luminosity of each CG requires the information of the CG members that have not been observed. Alternatively, the total luminosity  could be estimated and corrected from the total luminosity of the observed members based on the conditional luminosity of group members (e.g., \citealt{2007ApJ...671..153Y}, which will be preformed in the next work of our studies on the CGs.  Nevertheless, we emphasize that the total luminosity of the observed members only is already a good proxy of the total luminosity of the CGs since the undetected CG members contribute a small fraction of the total luminosity of the CGs. To confirm this conclusion, we show the total luminosity of the observed CG members as a function of the LOS velocity dispersion for the cCGs in different redshift bins in the bottom panel of figure~\ref{fig:flim}. In contrast to the richness, these scaling relations show few biases in different redshift bins.  

On the other hand, the missing of the bright galaxies ($r < 14.00 $) in our selection criteria might introduce  bias in the total luminosity of the group members. To validate this effect, we search the galaxies with $r < 14.00$ within 3$\theta_{G}$ for each cCG. We find 76 $r < 14.00$ galaxies associating with 74 cCGs.  Although our CGs are strictly defined on the galaxy sample with $14.00 \leq r \leq 17.77$, the group members and the total  luminosity of these CGs with very bright  association should be used with caution. To make compensate,  we list these $r < 14.00$ bright galaxies in appendix \ref{sec:extraBright}. As listed, most of them are located at $z < 0.03$.

Another possible inhomogeneity of our CG sample is the spectroscopic redshift source. As we have introduced in Section~\ref{sec:mgs}, the spectroscopic redshifts of our galaxy sample are heterogeneous. If there were systematical differences among different redshift catalogs, the LOS velocity dispersion measured for these CGs with heterogeneous redshifts would be biased to higher values. However, considering the typical uncertainty of the velocity measurements of current spectroscopy surveys ($\Delta V < 10$ km s$^{-1}$), and we also have tested,  this bias is negligible for our $\sigma_{\text{LOS}}$ measurements (see also \citealt{2016RAA....16...43S}).

\begin{figure}
\begin{center}
	\includegraphics[width=\columnwidth,keepaspectratio]{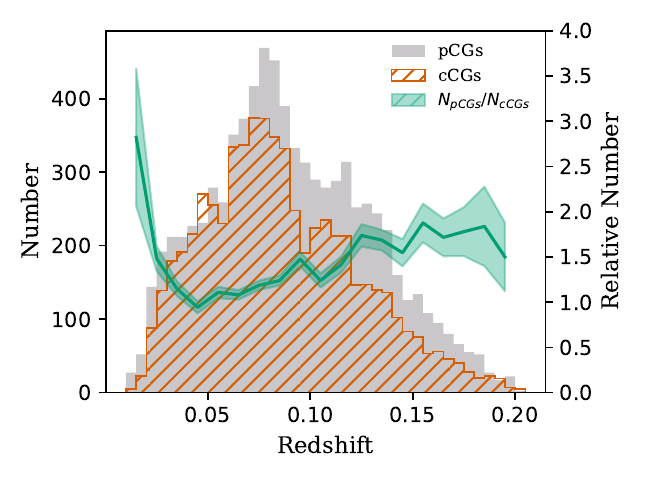}
    \caption{Redshift distributions of cCGs (red hatched) and pCGs (gray shaded). The green curve represents the ratio of the number of pCGs and cCGs at different redshift bins, the light region represents the 1$\sigma$ deviation from 1000 bootstraps. \label{fig:zcp}}
\end{center}
\end{figure}

\begin{figure}
\begin{center}
	\includegraphics[width=\columnwidth,keepaspectratio]{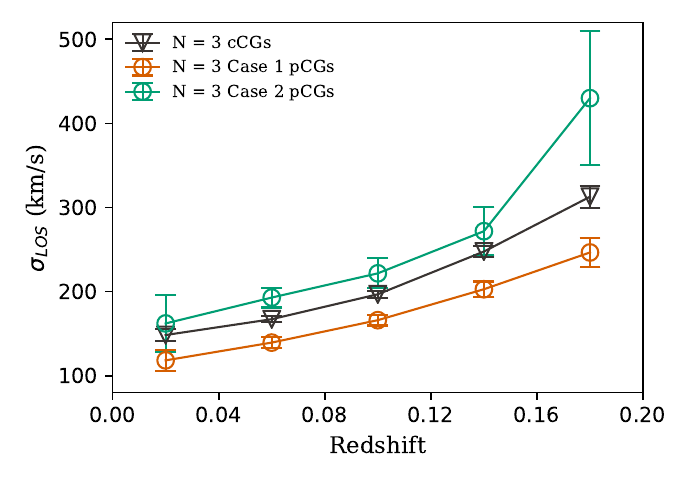}
    \caption{Median LOS velocity dispersion ($\sigma_{\text{LOS}}$) as a function of group redshift.
    The triangles, squares, and open circles represent $N = 3$ cCGs, $N = 3$ Case 1 pCGs, and $N = 3$ Case 2 pCGs respectively. \label{fig:psep}}
\end{center}
\end{figure}

\begin{deluxetable*}{lcccccc}
\tablecaption{Basic parameters of the member galaxies of each cCG, which include group ID, member ID, sky coordinates, redshift, $r$-band Magnitude and the source of Redshift Data. \tablenotemark{} \label{tab:cCGs-m}}
\tablewidth{2\columnwidth}
\tablehead{
\colhead{Group ID} & \colhead{Member ID} & \colhead{R.A.} & \colhead{Decl.} & \colhead{z} & \colhead{rmag} & \colhead{Source\tablenotemark{*}} \\
\colhead{} & \colhead{} & \colhead{(J2000)} & \colhead{(J2000)} & \colhead{} & \colhead{} & \colhead{}
}
%\colnumbers
\startdata
cCGs-0001 & 1 & 56.1398 & 1.0471 & 0.10312 & 17.565 & 1 \\
cCGs-0001 & 2 & 56.1366 & 1.0644 & 0.10349 & 16.929 & 1 \\
cCGs-0001 & 3 & 56.1375 & 1.0544 & 0.10615 & 16.133 & 1 \\
cCGs-0002 & 1 & 57.2892 & 0.8911 & 0.10910 & 16.755 & 1 \\
cCGs-0002 & 2 & 57.2610 & 0.9085 & 0.10912 & 17.428 & 2 \\
cCGs-0002 & 3 & 57.2689 & 0.8993 & 0.10953 & 17.040 & 1 \\
cCGs-0003 & 1 & 239.8613 & -0.9393 & 0.10117 & 17.484 & 1 \\
cCGs-0003 & 2 & 239.8510 & -0.9508 & 0.10268 & 16.262 & 1 \\
cCGs-0003 & 3 & 239.8504 & -0.9445 & 0.10328 & 17.268 & 1 \\
cCGs-0004 & 1 & 169.7703 & -0.2127 & 0.09620 & 16.237 & 4 \\
cCGs-0004 & 2 & 169.7850 & -0.2138 & 0.09750 & 17.506 & 1 \\
cCGs-0004 & 3 & 169.7718 & -0.2059 & 0.09867 & 17.498 & 1 \\
cCGs-0005 & 1 & 241.2199 & -0.0385 & 0.05027 & 16.202 & 1 \\
cCGs-0005 & 2 & 241.2295 & -0.0818 & 0.05055 & 16.848 & 1 \\
cCGs-0005 & 3 & 241.2225 & -0.0475 & 0.05197 & 15.854 & 1 \\
cCGs-0006 & 1 & 246.1209 & 0.7930 & 0.05699 & 17.100 & 1 \\
cCGs-0006 & 2 & 246.0963 & 0.7914 & 0.05803 & 16.513 & 1 \\
cCGs-0006 & 3 & 246.0957 & 0.8223 & 0.06000 & 15.940 & 1 \\
cCGs-0007 & 1 & 174.9234 & -1.0833 & 0.07654 & 16.472 & 3 \\
cCGs-0007 & 2 & 174.9079 & -1.0969 & 0.07772 & 17.034 & 3 \\
cCGs-0007 & 3 & 174.9048 & -1.0872 & 0.07905 & 17.025 & 3 \\
cCGs-0008 & 1 & 176.5291 & -1.0937 & 0.11589 & 16.259 & 3 \\
cCGs-0008 & 2 & 176.5184 & -1.1038 & 0.11633 & 17.509 & 1 \\
cCGs-0008 & 3 & 176.5396 & -1.0812 & 0.11855 & 16.642 & 1 \\
cCGs-0008 & 4 & 176.5341 & -1.1208 & 0.11861 & 16.540 & 3 \\
cCGs-0008 & 5 & 176.5346 & -1.1194 & 0.11886 & 16.890 & 1 \\
cCGs-0009 & 1 & 177.1664 & -1.2022 & 0.10535 & 17.357 & 1 \\
cCGs-0009 & 2 & 177.1699 & -1.1946 & 0.10649 & 17.670 & 3 \\
cCGs-0009 & 3 & 177.1756 & -1.2021 & 0.10794 & 17.466 & 1 \\
cCGs-0010 & 1 & 177.2069 & -1.1696 & 0.10577 & 17.285 & 1 \\
cCGs-0010 & 2 & 177.2084 & -1.1829 & 0.10682 & 16.856 & 3 \\
cCGs-0010 & 3 & 177.2039 & -1.1856 & 0.10682 & 17.615 & 1 
\enddata
\tablenotetext{}{\textcolor{gunen}{Note.} The group information is listed in table~\ref{tab:cCGs}. This table has 19,465 rows, only the first 32 rows are shown in here (corresponding to the first 10 cCGs listed in table~\ref{tab:cCGs}).}
\tablenotetext{*}{Source of the galaxy redshift: `0' for no redshift measurement; `1' for SDSS; `2' for LAMOST; `3' for GAMA; `4' for VAGC others.}
%\tablecomments{}
\end{deluxetable*}

\begin{deluxetable*}{lcccccc}
\tablecaption{Basic parameters of member galaxies of each groups identified in pCGs, which include group ID, member ID, sky coordinates, redshift, $r$-band Magnitude and the source of Redshift Data.  \tablenotemark{} \label{tab:pCGs-m}}
\tablewidth{2\columnwidth}
\tablehead{
\colhead{Group ID} & \colhead{Member ID} & \colhead{R.A.} & \colhead{Decl.} & \colhead{z} & \colhead{rmag} & \colhead{Source} \\
\colhead{} & \colhead{} & \colhead{(J2000)} & \colhead{(J2000)} & \colhead{} & \colhead{} & \colhead{}
}
\startdata
pCGs-0001 & 1 & 241.3809 & -0.5255 & - & 17.569 & 0 \\
pCGs-0001 & 2 & 241.3634 & -0.5315 & 0.12959 & 17.338 & 1 \\
pCGs-0001 & 3 & 241.3788 & -0.5243 & 0.13099 & 16.963 & 1 \\
pCGs-0002 & -99\tablenotemark{*} & 242.1488 & -0.1411 & - & 17.764 & 0 \\
pCGs-0002 & 1 & 242.1480 & -0.1375 & 0.13994 & 17.304 & 1 \\
pCGs-0002 & 2 & 242.1566 & -0.1092 & 0.13991 & 17.518 & 1 \\
pCGs-0002 & 3 & 242.1590 & -0.1264 & 0.14023 & 17.028 & 1 \\
pCGs-0003 & 1 & 242.6924 & -0.1257 & - & 16.772 & 0 \\
pCGs-0003 & 2 & 242.6894 & -0.1167 & 0.10626 & 17.149 & 1 \\
pCGs-0003 & 3 & 242.6807 & -0.1436 & 0.10638 & 16.620 & 1 \\
pCGs-0004 & 1 & 239.8467 & 0.3686 & - & 15.845 & 0 \\
pCGs-0004 & 2 & 239.8125 & 0.3963 & 0.14048 & 17.345 & 1 \\
pCGs-0004 & 3 & 239.8170 & 0.3658 & - & 16.776 & 0 \\
pCGs-0004 & 4 & 239.8282 & 0.3653 & 0.13932 & 17.554 & 1 \\
pCGs-0005 & 1 & 239.8170 & 0.3658 & - & 16.776 & 0 \\
pCGs-0005 & 2 & 239.8523 & 0.3621 & 0.09352 & 16.894 & 1 \\
pCGs-0005 & 3 & 239.8467 & 0.3686 & - & 15.845 & 0 \\
pCGs-0005 & 4 & 239.8157 & 0.3699 & 0.09331 & 17.073 & 1 \\
pCGs-0006 & -99\tablenotemark{*} & 242.6693 & 0.3061 & - & 17.024 & 0 \\
pCGs-0006 & 1 & 242.6529 & 0.3126 & 0.08022 & 16.965 & 1 \\
pCGs-0006 & 2 & 242.6644 & 0.3029 & - & 16.778 & 0 \\
pCGs-0006 & 3 & 242.6456 & 0.3017 & - & 17.138 & 0 \\
pCGs-0007 & 1 & 242.6629 & 0.2957 & 0.05806 & 15.694 & 1 \\
pCGs-0007 & 2 & 242.6644 & 0.3029 & - & 16.778 & 0 \\
pCGs-0007 & 3 & 242.6693 & 0.3061 & - & 17.024 & 0 \\
pCGs-0008 & 1 & 242.7062 & 0.3129 & 0.10622 & 15.816 & 1 \\
pCGs-0008 & 2 & 242.7307 & 0.3150 & - & 17.304 & 0 \\
pCGs-0008 & 3 & 242.7080 & 0.3187 & - & 17.274 & 0 \\
pCGs-0009 & 1 & 242.5368 & 0.7697 & 0.04192 & 14.900 & 1 \\
pCGs-0009 & 2 & 242.5338 & 0.7707 & - & 17.295 & 0 \\
pCGs-0009 & 3 & 242.5378 & 0.7652 & - & 14.459 & 0 \\
pCGs-0010 & 1 & 243.3677 & 0.8186 & 0.08188 & 17.370 & 1 \\
pCGs-0010 & 2 & 243.3666 & 0.8184 & - & 17.344 & 0 \\
pCGs-0010 & 3 & 243.3797 & 0.8135 & 0.08204 & 17.586 & 1 
\enddata
\tablenotetext{}{\textcolor{gunen}{Note.} The corresponding group information is listed in table~\ref{tab:pCGs}. This table has 26,652 rows, 25,132 rows correspond to the members of each pCGs and the rest 1,520 rows list the `interlopers' of Case 2 \& 3 pCGs (member ID: -99). Here we list the first 34 rows (corresponding to the first 10 pCGs listed in table~\ref{tab:pCGs}).}
\tablenotetext{*}{`-99' for the `interlopers' of Case 2 \& 3 pCGs.}
\end{deluxetable*}

\subsection{cCGs versus pCGs} \label{sec:discuss-pcgs}
In this study, we have also obtained a sample of 8,022 pCGs that have not been discussed yet. We show  the redshift distributions of cCGs and pCGs as hatched and shaded histograms in Figure~\ref{fig:zcp} respectively. For clarity, we show the number ratio of  pCGs to cCGs as the green curve in Figure ~\ref{fig:zcp}. This ratio shows a minimum at $z \sim 0.05$, and increases steeply and slowly toward the low and high redshift ends. This trend is to be expected: the nearby pCGs have larger radius and are more likely to be contaminated by the background galaxies' chance to be lie within 3$\theta_{G}$; while for  distant CGs, their angular radii are smaller on average, where the fiber collision effect becomes more significant and makes cCGs less complete.

We have also measured $\sigma_{\text{LOS}}$ for pCGs, where the member galaxies without spectroscopic redshifts are simply masked. Since not all pCGs fulfill with the criteria of cCGs, we would expect that $\sigma_{\text{LOS}}$ of pCGS would be biased from that of cCGs. We show such  a plot in Fig.\ref{fig:psep}, where the average $\sigma_{\text{LOS}}$  of $N = 3$ cCGs, Case 1 pCGs and Case 2 pCGs are plotted as functions of redshift. As expected, because of the interloper effect, at a given redshift, Case 1 pCGs have lower $\sigma_{\text{LOS}}$ while Case 2 pCGs have larger $\sigma_{\text{LOS}}$ than cCGs.

Ideally, if the redshift measurements of the SDSS MSGs are completed in the future, all our CG candidates, i.e. pCGs, could be redetermined as either real CGs (i.e. cCGs) or contaminators. Therefore, it is informative to have a estimation on the fraction of pCGs that could be identified as cCGs. To do that, we mask all the redshifts taken from the LAMOST spectral survey and make the same CG identification flow chart again. In this case, we obtain 1770 additional pCGs. Among them, 1010 are now identified as  cCGs using LAMOST redshifts. Taking the fraction of pCGs that are cCGs ($50\%-60\%$, see also\citealt{2008MNRAS.387.1281M}), we estimate there are about $\sim 5000$ genuine CGs in our pCG catalog.

\begin{deluxetable*}{cccccc}
\tablecaption{Comparison between cCGs and other CG catalogues derived from the SDSS. \label{tab:summerize}}
\tablecolumns{6}
\tablewidth{2\columnwidth}
\tablehead{
\multicolumn{2}{c}{} &
\colhead{cCGs\tablenotemark{}} & \colhead{S15\tablenotemark{}} & \colhead{MLCGs\tablenotemark{}} & \colhead{HMCGs\tablenotemark{}}
}
\startdata
& SDSS Data Release & DR14+ & DR12+ & DR12+ & DR12+ \\
Galaxy & Magnitude System & Petro &  Petro & Model & Model \\
Sample & Bright End & 14.00 & 14.50 & - & - \\
& Faint End & 17.77 & 18.00 & 17.77 & 17.77 \\
\hline
& Richness & $3 \leq N \leq 10$  & $N \geq 3$ & $N \geq 3$ & $4 \leq N \leq 10$ \\
& Surface Brightness & $\mu \leq 26.00$ & $\mu \leq 26.00$ & $\mu \leq 26.00$ & $\mu \leq 26.33$ \\
& Isolation & $\theta_{n} \geq 3 \theta_{G}$ & $\theta_{n} \geq 3 \theta_{G}$ & Unlimited & $\theta_{n} \geq 3 \theta_{G}$ \\
Criteria & Projected Separation Limit  & Unlimited & Unlimited & $\leq 50 h^{-1}$kpc & Unlimited \\
& Radial Velocity Limit  & $\leq 1000$ km s$^{-1}$  & $\leq 1000$ km s$^{-1}$ & $\leq 1000$ km s$^{-1}$ & $\leq 1000$ km s$^{-1}$ \\
& Magnitude Limit & $14.00 \le r \le 17.77$ & $\Delta m \leq 3$ & $\Delta m \leq 3$ & $\Delta m \leq 3$ \\
& & & & & $r_b \leq 14.77$ \\
\hline
\multicolumn{2}{c}{Total Groups} & 6,144  & 332 & 1,588 & 462 
\enddata
\end{deluxetable*}

\section{Summary} \label{sec:summary}
In this paper, we present two catalogs of CGs identified from the SDSS main sample galaxies ($14.00 \leq r \leq 17.77$) supplied with a significant fraction of redshifts from alternative surveys (e.g. LAMOST spectral survey and GAMA). Our motivation is to take the advantages of additional redshifts and maximize the final CG sample for statistical studies in next. Similar to \citet{2018AA...618A.157D}, our CG selection algorithm combines the advantages of two traditional CG selection algorithms, the photometric Hickson Criteria  and spectroscopic FoF method, so as to avoid possible selection biases in either low or high redshifts. Our final genuine CG catalog (cCGs) contains 6,144 $N \geq 3$ groups with 19,465 member galaxies, and 8,022 CG candidates (pCGs) catalog with 25,132 members, which are the largest spectroscopic CG catalogs to date. We perform a detailed comparison of our CG catalog with other available CG catalogs (S15, MLCG, and HMCG). The difference and improvement of  our CG selection algorithm are mainly reflected in the following way:

\begin{enumerate}
\item We extend the richness criterion to include the galaxy triplets.
\item We select the CGs in redshift slices, which prevents low-$z$ CGs being rejected if using photometric galaxy sample only.
\item We do not require  $\Delta m \leq 3$ for CG  members, which significantly  enlarges the sample size. The resulting inhomogeneity of the CG sample could be corrected in future statistical studies.
\item We keep all possible CG candidates in our pCG catalog, which could  be identified with further new redshifts.
\end{enumerate}

With this large cCG sample and  new CGs replenished  from the pCGs with future new redshifts (e.g. from LAMOST complementary galaxy sample, \citealt{2016RAA....16...43S}), in our next work, we will perform a detailed statistical study on the physical nature of the CGs, e.g. dynamics, environment, and member galaxy properties etc.

\acknowledgments
This work is supported by  National Key R$\&$D Program of China (No. 2019YFA0405501), National Natural Science Foundation of China (No.11573050 and 11433003) and Cultivation Project for LAMOST Scientific Payoff and Research Achievement of CAMS-CAS.

This work has made use of data products from the Sloan Digital Sky Survey (SDSS, \url{http://www.sdss.org}), the Large Sky Area Multi-Object Fiber Spectroscopic Telescope (LAMOST, \url{http://www.lamost.org}), the GAMA survey (\url{http://www.gama-survey.org}. We are thankful for their tremendous efforts on the surveying work.

Guoshoujing Telescope (the Large Sky Area Multi-Object Fiber Spectroscopic Telescope LAMOST) is a National Major Scientific Project built by the Chinese Academy of Sciences. Funding for the project has been provided by the National Development and Reform Commission. LAMOST is operated and managed by the National Astronomical Observatories, Chinese Academy of Sciences.

%\clearpage
\begin{appendix}
\section{List of the bright galaxies within $3 \theta_{G}$ of each cCGs} \label{sec:extraBright}
In this appendix, we list the bright galaxies ($r < 14.00$) within $3 \theta_{G}$ of each cCGs in Table~\ref{tab:extraBright},  including sky coordinates, redshift, galactic extinction corrected $r$-band Petrosian magnitude, corresponding cCG ID, redshift of corresponding cCG and the projected location in corresponding cCG.

Since the bright galaxies in SDSS are easily contaminated by the deblending effect,  we have made visual inspection for all these galaxies. There are 76 bright galaxies associated with 74 cCGs. If we join these bright galaxies into the current cCG catalog (defined on galaxies with $14.00 \leq r \leq 17.77$), 26 cCGs are still identified as cCGs (but the richness and surface brightness would be raised), while the other 48 would be rejected.

\begin{deluxetable*}{ccccccc}
\tablewidth{2\columnwidth}
\tablecaption{Basic parameters of bright galaxies within $3 \theta_{G}$ of Each cCGs, which include sky coordinates, redshift, galactic extinction corrected $r$-band Petrosian magnitude, Corresponding cCG ID, Redshift of Corresponding cCG and the Projected Location in Corresponding cCG \tablenotemark{} \label{tab:extraBright}}
\tablehead{
\colhead{R.A.} & \colhead{Decl.} & \colhead{$z$} & \colhead{rmag} & \colhead{Group ID} & \colhead{$z_{G}$} & \colhead{Location\tablenotemark{*}} \\
\colhead{(J2000)} & \colhead{(J2000)} & \colhead{} & \colhead{} & \colhead{} & \colhead{} & \colhead{}
}
\startdata
31.3690 & 13.2516 & 0.02603 & 12.904 & cCGs-0226 & 0.02557 & 1 \\
24.7282 & 15.0216 & 0.02790 & 13.806 & cCGs-0236 & 0.02807 & 2 \\
131.4607 & 53.9925 & 0.03095 & 13.778 & cCGs-0462 & 0.03069 & 1 \\
121.3606 & 46.7078 & 0.02266 & 13.306 & cCGs-0503 & 0.02246 & 1 \\
212.8593 & 1.2865 & 0.02494 & 13.445 & cCGs-0571 & 0.02516 & 2 \\
216.1382 & 1.1773 & 0.03879 & 13.728 & cCGs-0573 & 0.03848 & 2 \\
230.0759 & 3.5183 & 0.03691 & 13.060 & cCGs-0730 & 0.03769 & 2 \\
27.2866 & 13.059 & 0.01743 & 13.963 & cCGs-0845 & 0.01725 & 1 \\
27.3085 & 13.0554 & 0.01693 & 12.488 & cCGs-0845 & 0.01725 & 1 \\
172.0738 & 2.6540 & 0.02282 & 13.799 & cCGs-0901 & 0.02263 & 1 \\
128.5946 & 48.0882 & 0.04294 & 13.906 & cCGs-0928 & 0.04341 & 1 \\
158.5600 & 61.6405 & 0.03114 & 13.576 & cCGs-0947 & 0.03061 & 2 \\
255.2513 & 39.5661 & 0.03426 & 13.719 & cCGs-1172 & 0.03417 & 1 \\
209.8532 & -3.2092 & 0.02448 & 13.399 & cCGs-1395 & 0.02475 & 1 \\
50.1789 & -1.1086 & 0.02091 & 13.890 & cCGs-1670 & 0.02105 & 2 \\
50.1892 & -1.0447 & 0.02140 & 13.508 & cCGs-1670 & 0.02105 & 2 \\
177.7127 & 55.1437 & 0.0193 & 13.531 & cCGs-1714 & 0.01898 & 2 \\
139.1951 & 43.7126 & 0.0311 & 13.901 & cCGs-1894 & 0.02833 & 2 \\
187.1314 & 53.5961 & 0.03635 & 13.963 & cCGs-1917 & 0.03597 & 2 \\
199.1243 & 52.9337 & 0.0327 & 13.811 & cCGs-2189 & 0.03262 & 1 \\
173.9015 & 54.9486 & 0.01929 & 12.863 & cCGs-2225 & 0.01859 & 1 \\
45.9705 & 0.4152 & 0.04300 & 13.664 & cCGs-2364 & 0.04220 & 2 \\
155.6559 & 38.5791 & 0.05175 & 13.992 & cCGs-2626 & 0.05091 & 2 \\
145.8298 & 36.2478 & 0.02234 & 13.182 & cCGs-2636 & 0.02149 & 1 \\
228.3072 & 40.5379 & 0.03124 & 13.740 & cCGs-2658 & 0.03099 & 1 \\
218.2908 & 53.2328 & 0.04416 & 13.578 & cCGs-2836 & 0.04399 & 1 \\
187.5494 & 47.0063 & 0.03910 & 13.614 & cCGs-2843 & 0.03989 & 2 \\
184.5536 & 44.1733 & 0.02453 & 13.664 & cCGs-2954 & 0.02474 & 1 \\
198.1661 & 12.5998 & 0.01121 & 12.876 & cCGs-3099 & 0.01135 & 1 \\
203.1773 & 7.3273 & 0.02338 & 13.030 & cCGs-3197 & 0.02372 & 2 \\
207.0208 & 7.3923 & 0.02328 & 13.903 & cCGs-3208 & 0.02313 & 2 \\
235.1470 & 28.3607 & 0.03271 & 13.806 & cCGs-3347 & 0.03063 & 1 \\
219.5425 & 9.3361 & 0.03029 & 13.678 & cCGs-3435 & 0.03086 & 2 \\
204.0058 & 6.5853 & 0.02195 & 13.634 & cCGs-3437 & 0.02222 & 2 \\
223.1732 & 7.9319 & 0.03552 & 13.664 & cCGs-3453 & 0.03546 & 2 \\
234.2215 & 4.7579 & 0.03909 & 13.97 & cCGs-3489 & 0.03907 & 2 \\
171.8910 & 36.0610 & 0.03462 & 13.862 & cCGs-4194 & 0.03448 & 2 \\
141.9700 & 29.9857 & 0.02666 & 13.130 & cCGs-4351 & 0.02627 & 2 \\
205.2964 & 30.3781 & 0.04039 & 13.888 & cCGs-4537 & 0.04019 & 1 \\
196.3090 & 31.9997 & 0.05188 & 13.912 & cCGs-4538 & 0.05183 & 2 \\
239.8022 & 20.7634 & 0.01466 & 13.605 & cCGs-4649 & 0.01416 & 2 \\
194.2975 & 29.0451 & 0.02496 & 13.749 & cCGs-4660 & 0.02512 & 2 \\
224.6263 & 23.9553 & - & 13.986 & cCGs-4673 & 0.04781 & 1 \\
241.5951 & 19.7780 & 0.03911 & 13.845 & cCGs-4802 & 0.03901 & 1 \\
219.0386 & 21.7935 & 0.01877 & 12.907 & cCGs-4866 & 0.01750 & 2 \\
236.2438 & 16.9621 & 0.04953 & 13.974 & cCGs-4869 & 0.04920 & 2 \\
226.7999 & 20.4796 & 0.04215 & 13.725 & cCGs-4878 & 0.04118 & 2 \\
167.6600 & 28.7676 & 0.03479 & 13.428 & cCGs-5058 & 0.03703 & 2 \\
184.9485 & 30.3391 & 0.02817 & 13.039 & cCGs-5084 & 0.02772 & 2 \\
139.2788 & 20.0697 & 0.02784 & 12.891 & cCGs-5116 & 0.02922 & 2 \\
195.4739 & 27.6244 & 0.02621 & 12.940 & cCGs-5166 & 0.02334 & 2 \\
181.5246 & 28.2380 & 0.02734 & 13.906 & cCGs-5171 & 0.02905 & 2 \\
168.0139 & 27.5898 & 0.03500 & 13.438 & cCGs-5230 & 0.03482 & 2 \\
188.9216 & 26.5231 & 0.02221 & 12.231 & cCGs-5232 & 0.02298 & 2 \\
194.1161 & 26.9874 & 0.02150 & 13.023 & cCGs-5239 & 0.02097 & 2 \\
194.8988 & 27.9593 & 0.02391 & 12.408 & cCGs-5248 & 0.02343 & 2 \\
164.6051 & 24.2263 & 0.02145 & 13.090 & cCGs-5312 & 0.02131 & 1 \\
162.1901 & 22.2178 & 0.04667 & 13.845 & cCGs-5320 & 0.04428 & 2 \\
170.6098 & 24.2991 & 0.02515 & 13.241 & cCGs-5330 & 0.02514 & 1 \\
144.7567 & 17.0253 & 0.02986 & 13.953 & cCGs-5371 & 0.02929 & 2 \\
174.4318 & 22.0098 & 0.03009 & 13.764 & cCGs-5389 & 0.03057 & 2 \\
213.2858 & 20.4163 & 0.01672 & 12.299 & cCGs-5414 & 0.01619 & 1 \\
120.5625 & 9.3944 & 0.01490 & 13.538 & cCGs-5457 & 0.01437 & 2 \\
167.4352 & 21.7589 & 0.03183 & 13.316 & cCGs-5492 & 0.03165 & 2 \\
175.6020 & 20.1193 & 0.01996 & 13.810 & cCGs-5593 & 0.02091 & 2 \\
176.1273 & 20.0767 & 0.02241 & 13.789 & cCGs-5594 & 0.02429 & 2 \\
181.0060 & 20.2323 & 0.02445 & 12.970 & cCGs-5595 & 0.02217 & 2 \\
181.0391 & 20.3479 & 0.02454 & 12.818 & cCGs-5596 & 0.02410 & 2 \\
174.6228 & 20.5277 & 0.02572 & 13.358 & cCGs-5603 & 0.02532 & 1 \\
227.0999 & 19.2081 & 0.02109 & 13.780 & cCGs-5613 & 0.02113 & 2 
\enddata
%\tablecomments{}
\end{deluxetable*}

\setcounter{table}{7}
\begin{deluxetable*}{ccccccc}
\tablewidth{2\columnwidth}
\tablecaption{(Continued) \label{tab:extraBright-continue}}
\tablehead{
\colhead{R.A.} & \colhead{Decl.} & \colhead{$z$} & \colhead{rmag} & \colhead{Group ID} & \colhead{$z_{G}$} & \colhead{Location\tablenotemark{*}} \\
\colhead{(J2000)} & \colhead{(J2000)} & \colhead{} & \colhead{} & \colhead{} & \colhead{} & \colhead{}
}
\startdata
243.2081 & 11.1598 & 0.04247 & 13.675 & cCGs-5653 & 0.04118 & 2 \\
199.3690 & 20.6123 & 0.02222 & 13.664 & cCGs-5666 & 0.02265 & 1 \\
213.0659 & 15.8419 & 0.01750 & 13.232 & cCGs-5786 & 0.01742 & 1 \\
221.8119 & 13.4564 & 0.02802 & 13.909 & cCGs-5806 & 0.02831 & 2 \\
209.7617 & 15.5657 & 0.02559 & 13.135 & cCGs-5808 & 0.02513 & 1 \\
135.7972 & 13.6323 & 0.02903 & 13.554 & cCGs-5914 & 0.02913 & 2 
\enddata
\tablenotetext{}{\textcolor{gunen}{Note.} This table has 76 rows.}
\tablenotetext{*}{Location of the bright galaxies in their corresponding cCGs: `1' for `within the smallest enclosed circle'; `2' for `within the isolation ring'}
%\tablecomments{}
\end{deluxetable*}

\section{Example SDSS Images of CGs and CG candidates} \label{sec:example}
\begin{figure*}
\begin{center}
	\includegraphics[width=.9\textwidth,height=.9\textheight,keepaspectratio]{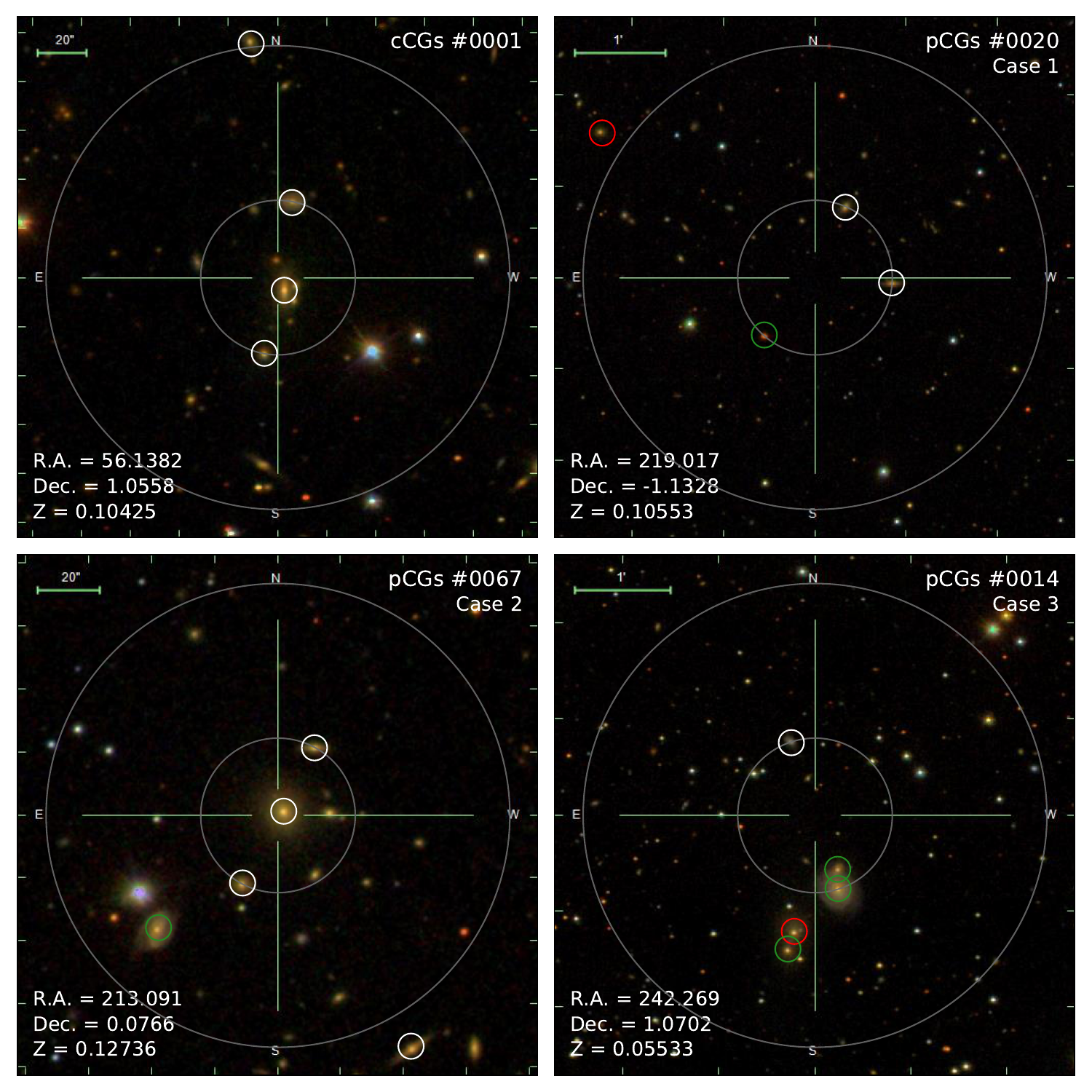}
    \caption{Example SDSS images of a cCG and each case of pCGs. For each CG image, the inner gray circle shows the smallest enclosed circles of the group members ($\theta_G$), while outer gray circle represents the corresponding isolation ring ($\sim 3\theta_G$). The small white circles locate the group members  with spectroscopic redshifts, while the small green circles label the galaxies without redshifts. The red circles represent the foreground or background galaxies. Only the galaxies in the magnitude range $14.00 < r < 17.77$ are labeled. \label{fig:ccgs-samples}}
\end{center}
\end{figure*}

\begin{figure}
\begin{center}
	\includegraphics[width=\columnwidth,height=\columnwidth,keepaspectratio]{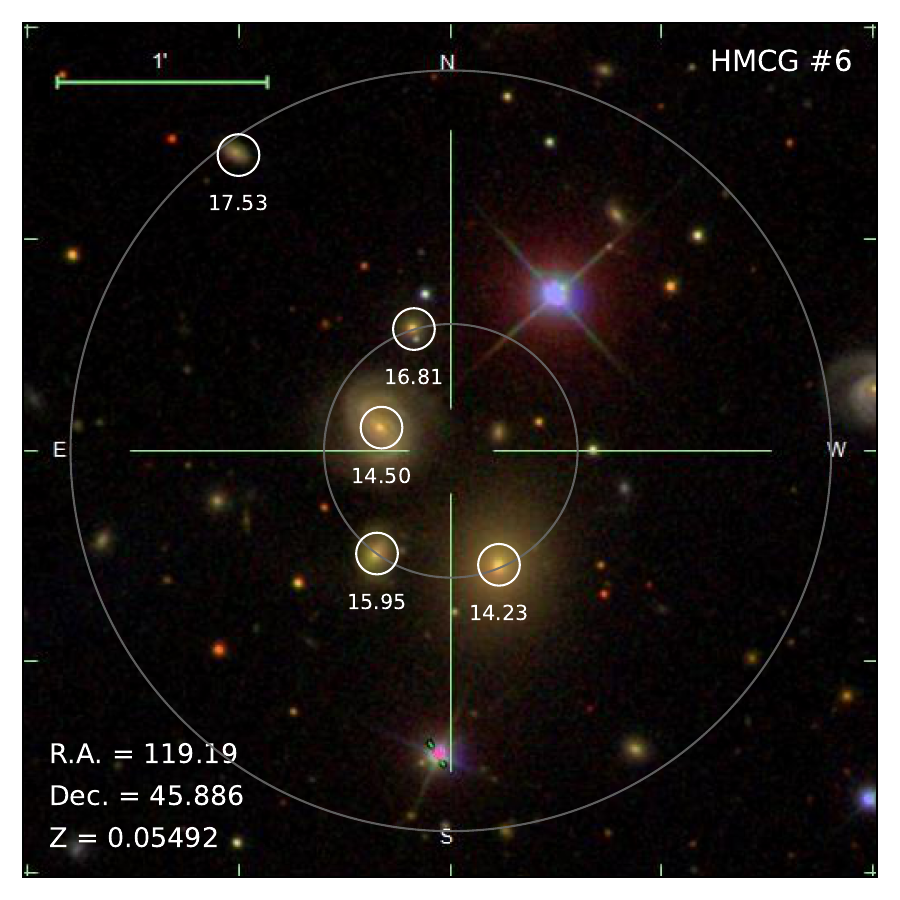}
    \caption{Example SDSS image of an HMCG that is not identified as a cCG. The symbol styles are the same as Fig.\ref{fig:ccgs-samples}, where the $r$-band magnitudes of each galaxies are also labelled. Because of the galaxy with $r = 17.53$ located inside $3\theta_G$, which is more than 3 magnitude fainter than the brightest member ($r = 14.23$) and has not been considered as a group member in HMCG, this group is not identified as a CG in our study. \label{fig:range3}}
\end{center}
\end{figure}

\noindent Example SDSS images: Fig.~\ref{fig:ccgs-samples} shows the images of different cases of cCGs and pCGs, Fig.~\ref{fig:range3} shows an example HMCG not identified as a cCG.

\end{appendix}

\end{CJK*}
\end{document}